\documentclass[a4paper, amsfonts, amssymb, amsmath, reprint, showkeys, nofootinbib, twoside, superscriptaddress]{revtex4-1}
\usepackage[english]{babel}
\usepackage[utf8]{inputenc}
\usepackage{bm}
\usepackage[colorinlistoftodos, color=green!40, prependcaption]{todonotes}
\usepackage{amsthm}
\usepackage{mathtools}
\usepackage{physics}
\usepackage{xcolor}
\usepackage{graphicx}
\usepackage[left=23mm,right=13mm,top=35mm,columnsep=15pt]{geometry} 
\usepackage{adjustbox}
\usepackage{placeins}
\usepackage[T1]{fontenc}
\usepackage{lipsum}
\usepackage{csquotes}
\usepackage{siunitx}
\usepackage{caption}
\usepackage{varioref}
\usepackage{subcaption}
\usepackage[pdftex, pdftitle={Article}, pdfauthor={Author}, colorlinks=true,colorlinks,citecolor=blue,urlcolor=magenta,linkcolor=blue]{hyperref}
\labelformat{section}{Section #1} 
\labelformat{subsection}{Section #1} 
\labelformat{subsubsection}{Section #1}
\labelformat{subsubsubsection}{Section #1}
\labelformat{equation}{Eq.~(#1)} 
\labelformat{figure}{Fig.~#1} 
\labelformat{subfigure}{Fig.~\thefigure#1} 
\labelformat{table}{Tab.~#1} 
\labelformat{appendix}{Appendix #1}

\begin{document}
\title{Distinguishing Jordan and Einstein frames in gravity through entanglement}

\author{Sumanta Chakraborty}
\affiliation{School of Physical Sciences, Indian Association for the Cultivation of Science, Kolkata-700032, India}

\author{Anupam Mazumdar }
\affiliation{University of Groningen, 9747AG Groningen, the Netherlands}

\author{Ritapriya Pradhan}
\affiliation{School of Physical Sciences, Indian Association for the Cultivation of Science, Kolkata-700032, India}
\begin{abstract}

In general relativity, the use of conformal transformation is ubiquitous and leads to two different frames of reference, known as the Jordan and the Einstein frames. Typically, the transformation from the Jordan frame to the Einstein frame involves introducing an additional scalar degree of freedom, often already present in the theory. We will show that at the quantum level, owing to this extra scalar degree of freedom these two frames exhibit subtle differences that the entanglement between two massive objects can probe.

\end{abstract}
\maketitle
\textit{Introduction} --- As a competitor to Einstein's general theory of relativity, there exists a compelling theory, known as the Brans–Dicke theory of gravitation (sometimes also called the Jordan–Brans–Dicke theory)~\cite{PhysRev.124.925}. It is an example of a scalar-tensor theory of gravity, where the gravitational interaction is mediated by a scalar field as well as the tensor field. Intriguingly, the gravitational constant $G$ is no longer a fundamental constant but depends on the dynamics of a scalar field $\phi$. These two theories are designated by two frames of reference, known as the Einstein and the Jordan frames, and they are related by a special conformal transformation where the space-time metric is re-scaled by the scalar field, see~\cite{Sotiriou:2008rp} for a detailed review. The scalar-tensor theories of gravity have been studied widely as a generalization to Einstein's gravity in cosmology~\cite{DeFelice:2010aj}, resolving the big bang singularity, and of course in black hole physics~\cite{Biswas:2005qr, Biswas:2011ar}. Furthermore, such a class of theory can be considered as the low energy limit of string theory~\cite{Polchinski:1998rq} and appears in the Kaluza-Klein theory involving compactified extra spatial dimensions as well~\cite{Appelquist:1988fh}. The natural question arises: what happens to these frames at a quantum level? There are many studies that compare the two frames at a classical level, see~\cite{Faraoni:1999hp, Postma:2014vaa, Azri:2018gsz, Rondeau:2017xck, Chakraborty:2016ydo}. Here we wish to understand what happens at the quantum level, especially to some quantum observables leading to entanglement.

To settle this score, we observe that recently there has been a proposal to test the quantum nature of gravity in a lab via spin entanglement witness~\cite{Bose:2017nin, ICTS}, see also~\cite{Marletto:2017kzi}. The protocol is known as the quantum gravity-induced entanglement of mass (QGEM). This protocol intends to create macroscopic spatial quantum superposition in a lab with embedded spins to study the entanglement between two quantum systems {\it solely} via gravitational interaction. It is possible to screen/mitigate the long-range electromagnetic interaction by the construction of a Faraday cage in the experiment, see~\cite{vandeKamp:2020rqh, Schut:2023eux, Schut:2023hsy}. Also, there is an optomechanical protocol to test the spin-2 nature of the gravitational interaction by witnessing the entanglement between a quantum system and a photon mediated via graviton~\cite{Biswas:2022qto, Carney:2021vvt}, and time domain measurements to witness the entanglement~\cite{Hanif:2023fto}.

In brief, as long as we trust in an effective field theory approach to quantum gravity, where locality and causality are maintained, we would expect the two quantum massive objects to entangle even at the lowest order in the energy, i.e., at the level of Newtonian potential, see~\cite{Marshman:2019sne, Bose:2022uxe, Danielson:2021egj, Christodoulou:2022mkf, Carney_2019, Carney:2021vvt}. This particular observation is a reminiscence to Bell's test for the entanglement~\cite{Hensen:2015ccp}, which can be observed even in the limit when $\hbar \rightarrow 0$, as was shown in the large angular momentum limit ~\cite{GISIN199215, Peres}. Therefore, quantum entanglement is considered a genuine quantum correlation that has no classical analog and is helpful to discern quantum vs. classical systems~\cite{schrodinger_1936, schrodinger_1935}. The entanglement of masses via the quantum gravitational interaction is also reminiscence to the so-called "Local Operation and Classical Communication" (LOCC) theorem, which states that it is not possible to entangle two quantum systems if they were not entangled, to begin with via classical channel~\cite{Bennett:1996gf}. This would mean that two test masses (quantum) would entangle in the presence of gravitational interaction \textit{if and only if} gravity obeys the rules of quantum mechanics \cite{Bose:2017nin, Marshman:2019sne, Bose:2022uxe}. As a natural consequence, gravitational interaction with matter, if treated classically, will not yield any entanglement whatsoever, as shown in ~\cite{Marshman:2019sne, Bose:2022uxe, Vinckers:2023grv}. 

Entanglement test to quantum nature of gravity has also been proposed to probe the weak equivalence principle~\cite{Bose:2022czr}, physics beyond the Standard Model, such as axions~\cite{Barker:2022mdz}, extra spacetime dimensions~\cite{Elahi:2023ozf}, and also gravitational theories with non-local gravitational interactions motivated from string theory~\cite{Vinckers:2023grv}. One particular observable for entanglement is the concurrence for the bipartite systems~\cite{wooters1}. In this paper, we will use the same to understand how the entanglement features can discern the properties of the two frames, and in this regard, we will define a new observable, which we call a relative concurrence difference, which we define as:
\begin{equation}\label{Rel-c}
\Delta\equiv \frac{C^{J}(\mathcal{G})-C^{E}(\mathcal{G}=0)}{C^{J}(\mathcal{G})+C^{E}(\mathcal{G}=0)}
\end{equation}
where $C^{E}(\mathcal{G}=0)$ denotes concurrence in the Einstein frame in the absence of a scalar field, and $C^{J}(\mathcal{G})$ is the concurrence described in the Jordan frame. Note that this definition is very similar to the E\"otv\"os parameter, introduced in~\cite{Bose:2022uxe}. We will show that even if the scalar field mass is small, e.g. $m_\phi \rightarrow 0$, the relative concurrence can exhibit the difference between the two frames.

\textit{Connecting Einstein and Jordan frames} --- The two frames\footnote{In the Jordan frame all the geometrical quantities are denoted by $\widetilde\eta_{\mu\nu},\widetilde h_{\mu\nu},\cdots$, while in the Einstein frame they are just denoted by $\eta_{\mu\nu},h_{\mu\nu},\cdots$. In our convention, the flat Minkowski metric in Cartesian coordinate read $\eta_{\mu \nu}=\textrm{diag.}\left(-,+,+,+\right)$.} are related by a conformal transformation: $g_{\mu \nu}\rightarrow \widetilde{g}_{\mu \nu}=\Omega^2(x)g_{\mu \nu}$, where the conformal factor can be expressed as $\Omega^2(x)=1+\phi(x)$, i.e., we assume the linear dilaton profile for $\Omega(x)$. Thus, the metric in the Jordan frame can be expanded around the flat background and hence expressed as,
\begin{align}\label{5}
\widetilde{g}_{\mu \nu}&=\Omega^2 g_{\mu \nu}=(1+\phi)\left(\eta_{\mu \nu}+h_{\mu \nu}\right)
\nonumber
\\
&=\eta_{\mu \nu}+\left(h_{\mu \nu}+\phi \eta_{\mu \nu}\right)~,
\end{align}
where $h_{\mu\nu}$ is the metric perturbation around the flat background, which we will quantize. Therefore, the situation in the Jordan frame appears to be identical to that in the Einstein frame, with simply the perturbed metric $h_{\mu \nu}$, being replaced by $\widetilde{h}_{\mu\nu}\equiv h_{\mu \nu}+\phi \eta_{\mu \nu}$. We would like to point out that $\phi$ and $h_{\mu \nu}$ are of the same order in the perturbative expansion around the flat background. Therefore, the gravitational Lagrangian will now involve non-trivial interactions between the metric perturbation $h_{\mu \nu}$ and the scalar degree of freedom $\phi$, while the interaction Hamiltonian between gravity and matter in the Jordan frame will become:
\begin{align}
\widetilde{H}_{\mathrm{int}}&=-\frac{1}{2}\int d^3 \mathbf{x}\, \widetilde{T}_{\mu \nu} (t,\mathbf{x}) \widetilde{h}^{\mu \nu}(t,\mathbf{x})~,\nonumber\\
&=H_{\rm int}-\frac{1}{2}\int d^3 \mathbf{x} \; \phi(t,\mathbf{x})T(t,\mathbf{x})\,,
\label{6}
\end{align}
where $\widetilde{T}_{\mu \nu}$ is the energy-momentum tensor of the matter field in the Jordan frame and $H_{\rm int}$ is the interaction Hamiltonian in the Einstein frame (As another demonstration of the above difference we have presented the corresponding situation in the $f(R)$ theories of gravity in \ref{App_fR}). Note that, $\widetilde{T}_{\mu \nu}=T_{\mu \nu}+\textrm{terms~of~}\mathcal{O}(\phi)$. Thus in the interaction term, we can simply use the following replacement: $\widetilde{T}_{\mu \nu }(\mathbf{x})=T_{\mu \nu}(\mathbf{x})$, since the additional term of $\mathcal{O}(\phi)$ will contribute only at the second order. Thus in the Jordan frame, there is an additional contribution due to the scalar degree of freedom, which is $\phi(x)T(x)$\footnote{This extra term will create a difference between the two frames in the quantum gravitational entanglement between two masses at the linear order. This is also consistent with \cite{Ohta:2017trn}, where minimal couplings are shown to violate equivalence between the two frames, which is the case in linear order.}. Here $T\equiv \eta^{\mu \nu}T_{\mu \nu}$, is the trace of the matter energy-momentum tensor. Interestingly, the term $H_{\rm int}$ arises solely from the gravitational interaction with matter in the Einstein frame, denoted by $H_{\rm g}$, while the scalar contribution is denoted by $H_{\phi}$. Owing to the linear nature of quantum theory (at least in the weak field regime of gravity that we are interested in), both of these terms, namely $H_{\rm g}$ and $H_{\phi}$ can be studied separately. The contribution of $H_{\rm g}$ to the entanglement between two massive objects has already been computed \cite{Bose:2017nin, Marshman:2019sne, Bose:2022uxe, Vinckers:2023grv}, while in this paper, we will focus on the entanglement due to $H_{\phi}$. 

To proceed further, we need explicit expressions for the matter energy-momentum tensor. We assume that the two masses (quantum) are separated by a distance $d$ and they can move in one dimension only, so that the energy-momentum tensor reads,
\begin{equation}\label{11}
T_{\mu \nu}(\mathbf{r})=\frac{p_\mu p_\nu}{E/c^2}\Big[\delta(\mathbf{r}-\mathbf{r}_A)+\delta(\mathbf{r}-\mathbf{r}_B)\Big]~,
\end{equation}
where $p_{\mu}=(-E/c,\bm{p})$, $E=\sqrt{|\bm{p}|^{2}c^{2}+m^{2}c^{4}}$, with $\mu,\nu=0,1$, and $\bm{r}_{\rm A}=(x_{\rm A},0,0)$, and $\bm{r}_{\rm B}=(x_{\rm B},0,0)$ denoting the positions of the two masses. Thus, the only non-zero components of the energy-momentum tensor are $T_{00}$, $T_{01}$, and $T_{11}$.

{\it Interaction Hamiltonian in the Einstein frame} --- First, we recall the contribution of the interaction Hamiltonian $H_{\rm int}$ in the Einstein frame, which involves the tensor field alone and hence is also given by $H_{\rm g}$. By quantizing the spin-2 graviton in a similar fashion as~\cite{Marshman:2019sne, Bose:2022uxe, Vinckers:2023grv}, along with the quantum matter sector, we can compute the change in the gravitational energy $\Delta \hat H_{\rm g}$ of the system involving the two masses. As we assume the two masses and the graviton to be in their ground states, the above interaction excites the graviton and hence the masses as well. The excitation experienced by the masses is given by the interaction term $\Delta \hat H_{\rm g}$, which is given by:
\begin{equation}\label{eq:dHij}
\Delta\hat{H}_{g}\equiv\sum\int d\bm{k}\frac{\,_{\rm g}\langle0\vert\hat{H}_{\text{int}}\vert\bm{k}\rangle_{\rm g}\,_{\rm g}\langle\bm{k}\vert\hat{H}_{\text{int}}\vert0\rangle_{\rm g}}{E^{\rm g}_{0}-E^{\rm g}_{\bm{k}}}\,,
\end{equation}
where the above summation takes care of the polarization states of one particle graviton energy eigenstates $|\bm{k}\rangle_{\rm g}$, see~\cite{Bose:2022uxe, Vinckers:2023grv}. Moreover, $\vert0\rangle_{\rm g}$ denotes the vacuum state of the graviton and $E^{\rm g}$ is the energy of the graviton. By performing the above integration and keeping terms up to $\mathcal{O}(1/c^{4})$, the gravitational interaction experienced by the two (quantum) masses become \cite{Bose:2022uxe} (ignoring self-energy terms, see \ref{Appquant}): 
\begin{align}\label{eq:dH5}
\Delta\hat{H}_{g}&=-\frac{Gm^{2}}{\vert{\hat{\bf r}}_{A}-{\hat{\bf r}}_{B}\vert}
+\frac{4G\hat{p}_{A}\hat{p}_{B}}{c^{2}\vert{\hat{\bf r}}_{A}-{\hat{\bf r}}_{B}\vert}
-\frac{9G{\hat{p}}_{A}^{2}{\hat{p}}_{B}^{2}}{4c^{4}m^{2}\vert{\hat{\bf r}}_{A}-{\hat{\bf r}}_{B}\vert}\,.
\end{align}
The above expression for the change in the gravitational energy involving the two point masses matches with the known results from quantizing gravity at the linear level, see~\cite{Donoghue:1994dn, Grignani:2020ahv, Cristofoli:2019neg, Iwasaki:1971vb}. For perturbative quantum gravity, the change in the gravitational energy is an operator-valued entity, depending on the operators $\hat x_{\rm A}$, $\hat x_{\rm B}$, $\hat p_{\rm A}$, and $\hat p_{\rm B}$, rather than complex numbers~\cite{Bose:2022uxe, Vinckers:2023grv}. 

{\it Interaction Hamiltonian in the Jordan frame} --- The change in the gravitational energy of the two masses in the Jordan frame requires quantizing the dilaton/scalar field $\phi$. We assume that the scalar field has a mass $m_\phi$, and hence the canonical quantization scheme yields,
\begin{equation}
\phi = \chi \int d^3\mathbf{k}\;\sqrt{\frac{\hbar}{2\omega_\mathbf{k} (2\pi)^3}} \left(Q_{\mathbf{k}}^{\dagger}e^{-i\mathbf{k}\cdot \mathbf{r}}+Q_{\mathbf{k}}e^{i\mathbf{k}\cdot \mathbf{r}}\right)~,
\label{2}
\end{equation}
where the commutation rule between the creation operator $Q^\dagger_{\mathbf{k}}$ and the annihilation operator $Q_{\mathbf{k}}$ is given by, $[Q_{\mathbf{k}},Q^\dagger_{\mathbf{k}}]=\delta^{3}(\mathbf{k}-\mathbf{k'})$. Note that $\chi$ is an overall constant determining the strength of the scalar perturbation. Given the vacuum state of the scalar field $|0\rangle_{\phi}$, satisfying 
$Q_{\mathbf{k}}|0\rangle_{\phi}=0=\,_{\phi}\langle 0|Q_{\mathbf{k}}^\dagger$, 
it follows that the first order shift in Hamiltonian due to the interacting scalar field vanishes. Hence the shift in the interaction Hamiltonian of the two masses due to the exchange of a scalar field is given by, 
\begin{equation}
\Delta \hat{H}_{\phi}=\int d^3\mathbf{k'}\;\frac{\,_{\phi}\langle 0|\hat{H}^{\phi}_{\mathrm{int}}|\mathbf{k'}\rangle_{\phi} \,_{\phi}\langle \mathbf{k'}|\hat{H}^{\phi}_{\mathrm{int}}|0\rangle_{\phi}}{E^{\phi}_0 - E^{\phi}_{\mathbf{k'}}}~.
\label{5a}
\end{equation}
Here, we have defined the state $|\mathbf{k}\rangle_{\phi}$ to be the one-particle state of the $\phi$ field having momentum $\hbar\mathbf{k}$, i.e., $|\mathbf{k}\rangle_{\phi} \equiv Q^\dagger (\mathbf{k})|0\rangle$. By using the fact that $E^{\phi}_0-E^{\phi}_{\mathbf{k}}=-\hbar \omega_{\mathbf{k}}$, and ignoring the self-energy terms (which neither contribute to the infrared at low energies, not to the entanglement), in the non-relativistic regime\footnote{Non-relativistic in the sense that the (quantum) masses are ``moving slowly'', such that we can express the energy in terms of the momentum as, $E=mc^{2}\{1+(p^2/2m^2c^2)\}$, which also leads to $(mc^{2}/E)=\{1-(p^2/2m^2c^2)\}$.}, we obtain:
\begin{align}
\label{22a}
\Delta \hat H_{\phi}&=-\frac{e^{-|\mathbf{\hat r}_A-\mathbf{\hat r}_B|/L}}{|\mathbf{\hat r}_A-\mathbf{\hat r}_B|}\left(\mathcal{G}m^{2}+\frac{\mathcal{G}\hat p_A^2\hat p_B^2}{4m^2c^4}\right)~,
\end{align}
where, we have introduced a new length scale $L$ in the problem, defined as the inverse of the mass of the scalar field $m_{\phi}$, such that
\begin{equation}\label{19}
L\equiv\frac{\hbar}{m_\phi c}\,,~\mathcal{G}\equiv \frac{\chi^2 c^2}{16\pi}\,,
\end{equation}
and we have defined $\mathcal{G}$ as an effective strength for the scalar field. Therefore, the total interaction Hamiltonian for the two masses, due to the exchange of the scalar and the graviton degrees of freedom between them, can be written as (keeping terms up to second order in the post-Newtonian expansion and ignoring the self-energy terms),
\begin{align}\label{45}
\Delta \hat{H}_{\rm AB}&=\Delta \hat{H}_{g}+\Delta \hat{H}_{\phi} 
=-\frac{(G+\mathcal{G}e^{-({\bf \hat r_A-\hat r_B)}/L})m^2}{|\hat{\mathbf{r}}_A-\hat{\mathbf{r}}_B|}
\nonumber 
\\ 
&
+\frac{4G\hat{p}_A\hat{p}_B}{c^2 |\mathbf{\hat{r}}_A-\mathbf{\hat{r}}_B|}
-\frac{9(G+\frac{1}{9}\mathcal{G}e^{-({\bf \hat r_A-\hat r_B)}/L})\hat{p}_A^2\hat{p}_B^2}{4m^2c^4~|\mathbf{\hat{r}}_A-\mathbf{\hat{r}}_B|}~,
\end{align}
where we have explicitly shown the operators. From the above expression, \ref{45}, we can conclude the following\footnote{To the best of our knowledge the interaction Hamiltonian between the two masses in the Jordan frame has been computed for the first time in this paper}: The interaction Hamiltonian $\Delta \hat{H}_{\phi}$ has a term proportional to $m^2$ and a term proportional to $\hat p_{A}^{2}\hat p_{B}^{2}$, however the linear order term $\hat p_A\hat p_B$ is absent. In particular, the total effect of gravity and the scalar system can be encoded by defining an effective gravitational constant at the leading order:
\begin{equation}
\widetilde{G}=G+\mathcal{G}e^{-d/L}~,
\end{equation}
where, $d=|\bm{r}_{\rm A}-\bm{r}_{\rm B}|$. Note that in the limit $L\to 0$, or, $m_{\phi}\to \infty$, the modified gravitational constant in the Jordan frame becomes that of the Einstein frame, i.e., the same as Newton's constant $\widetilde G =G$.

On the other hand, at a higher order, namely $\mathcal{O}(1/c^4)$, the contribution of gravity and scalar adds up to yield a different effective gravitational constant, $\widetilde{G}'=G+(1/9)\mathcal{G}e^{-d/L}$. Thus it is interesting to observe that the effect of the scalar field is not a mere uniform translation of the gravitational constant at every order in $(1/c^{2})$, rather, it affects every order differently. 

\begin{figure}[ht]
\centering
\includegraphics[width=4.5cm,height=3.2cm,angle=0]{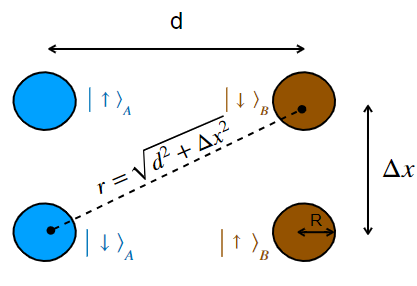}
\caption{Configuration of two massive objects separated by distance $d$ in spatial superposition, with the splitting being $\Delta x$. The radius of objects are $R\sim\left({3m}/{4\pi\rho}\right)^{1/3}$, so that $R\ll (\Delta x, d)$, for $\rho$=3.5 g $\textrm{cm}^{-3}$ typical of a diamond-like system with $m\sim10^{-14}$ kg.}
\label{fig:system configuration}
\end{figure}

{\it Concurrence in a QGEM setup} --- We will follow the original QGEM protocol~\cite{Bose:2017nin} to test the quantum nature of gravity via entanglement and will consider a parallel setup as discussed in \cite{Nguyen:2019huk, Tilly:2021qef, Schut:2021svd, Barker:2022mdz}. Such spatial superposition can be created in the Stern-Gerlach setup with a nitrogen valence (NV) spin embedded in the crystal~\cite{Bose:2017nin, Marshman:2021wyk, Margalit:2020qcy, Zhou:2022frl, PhysRevLett.125.023602, Marshman:2023nkh, Zhou:2022jug}. Here we mainly discuss the theoretical consequence of the concurrence in this setup and will place the masses in a superposition of size $\Delta x$ in a parallel arrangement, as shown in \ref{fig:system configuration}. We will now study how this spatial superposition/non-Gaussian state evolves under gravitational interaction Hamiltonian, with additional contribution from the scalar field. We start with the following initial state
$|\psi\rangle_{i}=\frac{1}{2}|C\rangle_A|C\rangle_B\left(|\uparrow\rangle_A+|\downarrow\rangle_A \right)\left(|\uparrow\rangle_B+|\downarrow\rangle_B \right)$,
where, $|C\rangle_A$ and $|C\rangle_B$ correspond to the localized state of particle A and particle B, respectively. After acting on the above state with appropriate stern-Gerlach apparatus, see~\cite{Margalit:2020qcy, PhysRevLett.125.023602, Marshman:2021wyk, Zhou:2022frl}, we arrive at the spatial superposition of the two test masses $A$ and $B$,
\begin{equation}\label{spin_superpose}
|\psi\rangle_{i} \rightarrow \frac{1}{2}\left(|L,\uparrow\rangle_A+|R,\downarrow\rangle_A \right)\left(|L,\uparrow\rangle_B+|R,\downarrow\rangle_B \right)~.
\end{equation}
The spatial superposition is created such that $|L\rangle_A$ and $|R\rangle_B$, as well as, $|R\rangle_A$ and $|L\rangle_B$, are separated by a distance $d$, while $|L\rangle_A$, $|L\rangle_B$ and $|R\rangle_A$, $|R\rangle_B$ are separated by a distance $\sqrt{d^{2}+(\Delta x)^{2}}$.
\FloatBarrier

In what follows, we will consider the non-relativistic limit and will take into account the leading order contributions, with $\hat p_A, \hat p_B \rightarrow 0$ in \ref{45}, such that the final state evolves to after time $t=\tau$,
\begin{align}
|\psi\rangle_{\rm f}&=e^{-i\Delta\hat{H}_{\rm AB}\tau}|\psi\rangle_{\rm i} 
\nonumber 
\\ 
&=\frac{1}{2}\left(e^{i\theta}|L,\uparrow\rangle_A|L,\uparrow\rangle_B+e^{i\phi}|L,\uparrow\rangle_A|R,\downarrow\rangle_B 
\nonumber \right. 
\\ 
&+\left. e^{i\phi}|R,\downarrow\rangle_A|L,\uparrow\rangle_B+e^{i\theta}|R,\downarrow\rangle_A|R,\downarrow\rangle_B \right)~,
\end{align}
where, two phases $\theta$ and $\phi$ are given by:
\begin{align}
\theta&=\left(\frac{Gm^2}{(d^2+\Delta x^2)^{1/2}}+\frac{\mathcal{G}e^{-(d^2+\Delta x^2)^{1/2}/L}m^2}{(d^2+\Delta x^2)^{1/2}}\right)\frac{\tau}{\hbar} ~,
\nonumber 
\\   
\phi&=\left(\frac{Gm^2}{d}+\frac{\mathcal{G}e^{-d/L}m^2}{d}\right)\frac{\tau}{\hbar} ~.
\end{align}
Subsequently, after passing through a full-loop interferometer, we obtain:
\begin{align}\label{final_wavef}
|\psi\rangle_{\rm f}&=\frac{e^{i\theta}}{2}|C\rangle_A|C\rangle_B\left[|\uparrow\rangle_A\left(|\uparrow\rangle_B+e^{i(\phi-\theta)}|\downarrow\rangle_B\right) 
\nonumber \right. 
\\ 
&+\left. |\downarrow\rangle_A\left(|\downarrow\rangle_B+e^{i(\phi-\theta)}|\uparrow\rangle_B\right)\right]~.
\end{align}
Subsequently, tracing out the spin degrees of freedom of $B$, we get the density matrix associated with the spin states of the mass $A$ as,
\begin{equation}
\hat{\rho}_A=\frac{1}{2}
\begin{pmatrix}
1&\cos(\phi-\theta)\\
\cos(\phi-\theta)&1
\end{pmatrix}~,
\end{equation}
and hence the following expression for the concurrence follows~\cite{wooters1}: 
\begin{equation}
C=\sqrt{2\left(1-\mathrm{tr}\left(\hat{\rho}_A^2\right)\right)}=|\sin(\phi-\theta)|~. 
\end{equation}

\begin{figure}
\centering
\includegraphics[width=8cm,height=5.7cm,angle=0]{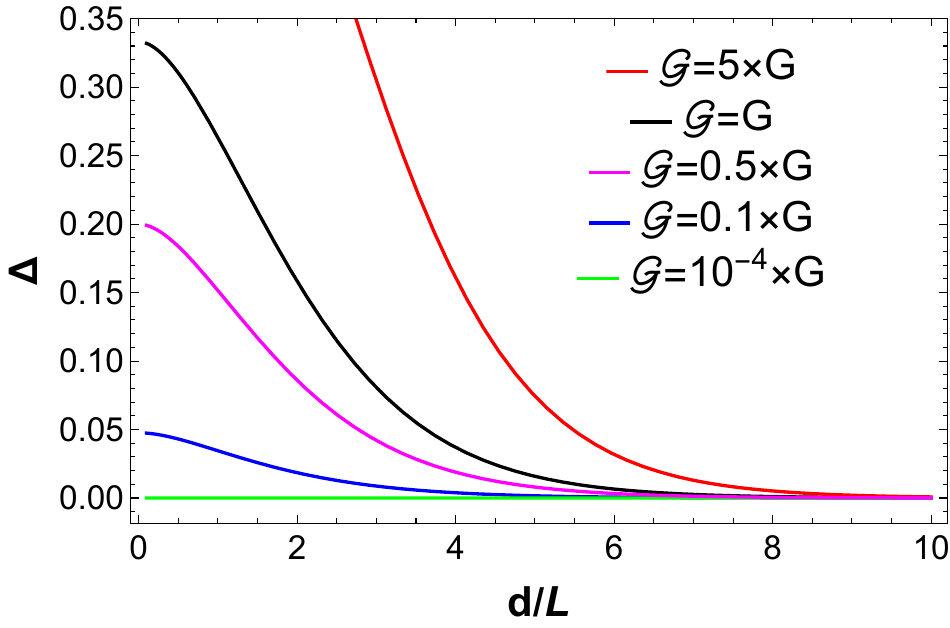}
\caption{Plot of relative concurrence  difference $\Delta$, see \ref{Rel-c}, with the dimensionless distance parameter $(d/L)$ has been presented, for different values of $(\mathcal{G}/G)$, where $G$ is the four-dimensional Newton's constant. We have set $d=200\mu$m, $\Delta x=100\mu$m, $m=10^{-14}$kg, and $\tau=1$s. For $L\rightarrow \infty$, implying $m_\phi\rightarrow 0$, we have the largest relative concurrence difference, implying the deviation between the Jordan and the Einstein frames is most pronounced there.}
\label{Figure-1}
\end{figure}

In \ref{Figure-1} we have illustrated how the relative concurrence difference $\Delta$ scales with $(d/L)$ for different values of the scalar coupling ${\cal G}$ in the Jordan frame. If the scalar coupling  ${\cal G}\rightarrow 0$, then $\widetilde G=G$, the Newton's constants in both the frames become identical. In this case, the relative concurrence difference tends to be zero, which can be verified from \ref{Figure-1}. We can see that the relative concurrence difference has the most pronounced effect in the limit of $L\rightarrow \infty$, and $m_\phi\rightarrow 0$, for any finite $d$. This very fact can be used as a probe to evidence what nature selects, the Jordan or the Einstein frame at a quantum level. In fact, this will allow us to probe light masses at an unprecedented level, which we had not been able to do so by any other gravitational experiment. 

We would like to emphasize that there are several constraints on the scalar-tensor theories arising from various classical experiments, ranging from solar system tests to gravitational wave measurements. However, all of these measurements probe the scalar field at a macroscopic level, while we have presented the effect of the scalar field at the quantum level, with a very distinct observable, with no classical analog. Moreover, the most stringent constraints from gravitational waves on some of the scalar-tensor theories, e.g., Horndeski \cite{Bettoni:2016mij}, can be lifted by advocating appropriate screening mechanisms and are likely to remain so even for future gravitational wave measurements \cite{Crisostomi:2017lbg}. On the other hand, at the quantum level, the existence of an additional scalar mediator results in entanglement, which can testify to the existence of a fifth force, in the cleanest possible way.

Indeed, there are various efforts to perform such kind of entanglement experiment on an atom chip, see~\cite{Margalit:2020qcy}, and there are many experimental challenges that need to be overcome. Creating superposition of such massive objects is another challenge, there might be the hypothetical collapse of the wavefunction~\cite{Penrose:1996cv, Diosi:1989}, which might even limit us to create such a large superposition, but such theories go beyond the scope of linearity of quantum mechanics while introducing non-linearities in gravity, hence we do not deal with such class of theories here. Besides these, there are known Standard Model interactions that can decohere the superposition, which severely puts a limit on the ambient temperature, the temperature of the object, pressure, impurities in the crystal, external gravitational stochastic  waves~\cite{Bose:2017nin, Schut:2021svd, Schut:2023hsy, vandeKamp:2020rqh, Toros:2020dbf, Toros:2020krn}.

Despite all these challenges, the import of this work is to show that at a quantum level the bonafide quantum observable, such as an entanglement/concurrence, exhibits the key difference between the Jordan and the Einstein frame. We have captured this by the relative concurrence difference $\Delta$, ~\ref{Rel-c}. The difference between the two frames arises from the propagation of an extra scalar degree of freedom in the Jordan frame. 
Though present generation experiments are not sensitive enough to detect the macroscopic superposition of massive objects, it is expected that within the next five to ten years this should be possible with the advancement of technology \cite{Marshman:2021wyk}. In which case, this feature will be testable in a laboratory setting involving the quantum superposition of two massive objects with a mass of $10^{-14}$~kg separated by a distance of $d=200 {\rm \mu m}$, and $\Delta x\sim 100{\rm \mu m}$, with sensitivity of $\mathcal{O}(10^{-1})$ for the relative concurrence. Therefore, quantum experiments are essential to unravel one of the mysteries of gravitational physics, in which frame the quantum observables make sense.
  
\begin{acknowledgments}

Research of S.C. is funded by the INSPIRE Faculty fellowship from the DST, Government of India (Reg. No. DST/INSPIRE/04/2018/000893). R.P. would like to thank Labani Roy for the useful discussions. Research of R.P. is supported by the IACS MS scholarship.  

\end{acknowledgments}
\appendix
\labelformat{section}{Appendix #1} 
\labelformat{subsection}{Appendix #1}

\section{Interaction Hamiltonian in f(R) theory of gravity}\label{App_fR}

As an example of the procedure outlined in the main text, we turn our attention to the $f(R)$ theory of gravity. It is well known that the $f(R)$ theory of gravity can be expressed as the Einstein gravity with an additional scalar degree of freedom and hence will be an ideal test bed to illustrate the crux of our argument \cite{DeFelice:2010aj, Sotiriou:2008rp, Nojiri:2010wj, Capozziello:2007ms, delaCruz-Dombriz:2015tye, delaCruz-Dombriz:2009pzc, Chakraborty:2015bja, Chakraborty:2015taq, Chakraborty:2016ydo, Chakraborty:2014xla, Chakraborty:2016gpg}. We demand that the flat metric $\eta _{\mu \nu}$ be a solution to the gravitational field equations in $f(R)$ theory of gravity, and hence the gravitational Lagrangian must depend only on the positive powers of the Ricci scalar. Thus we will only consider cases, where $f(R)=R+F(R)$, with $F(R)$ involving terms $\sim R^n$, with $n>1$. In this case, the interaction Hamiltonian is simply given by $-(1/2)h_{\mu \nu}T^{\mu \nu}$, however, the main question is, what equation do the gravitational perturbation $h_{\mu \nu}$ satisfy. In order to answer this question, we start by writing down the gravitational field equations for $f(R)$ gravity as,
\begin{align}
\label{9}
G_{\mu \nu}&+F'(R)R_{\mu \nu}-\frac{1}{2} F(R)g_{\mu \nu}
\nonumber
\\
&-\nabla_\mu \nabla_\nu F'(R)+g_{\mu \nu} \Box F'(R)=0~.
\end{align}
Expanding all the curvature quantities around the flat background in terms of the gravitational perturbation $h_{\mu \nu}$, it follows that, $F'(R) \sim F'(0)+F''(0)R+\mathcal{O}(h^2)$ and $F(R) \sim F(0)+F'(0)R+\mathcal{O}(h^2)$. Since $F(R)$ depends on positive powers of the scalar curvature and $F(R)=f(R)-R$, it follows that $F(0)=0=F'(0)$, while $F''(0)\neq 0$. Therefore, the linearized gravitational field equations reduce to,
\begin{equation}\label{n1}
G_{\mu \nu}-F''(0)\partial_\mu \partial_\nu R + \eta_{\mu \nu} F''(0) \Box R = 0~.
\end{equation}
Since we are working at the linear order in the perturbation, all the covariant derivatives have been converted to partial derivatives and the $\square$ operator is with respect to the flat background. Moreover, the trace of the above equation yields a simple wave-like equation for the Ricci scalar, $\Box R=\{1/3F''(0)\}R$, with an effective mass term $m_{R}=1/\sqrt{3F''(0)}$. All the geometrical quantities appearing in the above expression for the linearized gravitational field equation, namely $G_{\mu \nu}$, $R_{\mu \nu}$ and $R$ are also to be considered up to linear order in the perturbation, yielding, 
\begin{align}\label{12}
G_{\mu \nu}&=-\frac{1}{2}\partial_\mu \partial_\nu h-\frac{1}{2}\Box h_{\mu \nu}+\frac{1}{2}\partial_\nu \partial_\alpha h^\alpha_\mu  
\nonumber
\\
&\qquad +\frac{1}{2}\partial_\mu \partial_\alpha h^\alpha _\nu+\mathcal{O}(h^{2})~,
\\
R_{\mu \nu}&=-\frac{1}{2}\Box h_{\mu \nu}-\frac{1}{2}\partial_\mu \partial_\nu h + \frac{1}{2}\partial_\mu \partial_\alpha h^\alpha _\nu 
\nonumber
\\
&\qquad +\frac{1}{2}\partial_\nu \partial_\alpha h^\alpha _\mu+\mathcal{O}(h^{2})~,
\\
R&=-\Box h +\partial_\alpha \partial_\beta h^{\alpha \beta}+\mathcal{O}(h^{2})~.
\end{align}
Substituting these expressions in \ref{n1}, the linearized gravitational field equations, we obtain the following evolution equation for the gravitational perturbation $h_{\mu \nu}$,
\begin{align}\label{18'}
\frac{1}{2}&\partial_\nu \partial_\alpha h^\alpha_{\mu} + \frac{1}{2}\partial_\mu \partial_\alpha h^\alpha _\nu+\frac{1}{2} \eta_{\mu \nu} (\Box h-\partial_\alpha \partial_\beta h^{\alpha \beta}) 
\nonumber 
\\ 
& -\frac{1}{2}\Box h_{\mu \nu}
+F''(0)\partial_\mu \partial_\nu (\Box h-\partial_\alpha \partial_\beta h^{\alpha \beta})
\nonumber 
\\ 
&+\frac{1}{2}\partial_\mu \partial_\nu h-F''(0)\eta_{\mu \nu} \Box (\Box h-\partial_\alpha \partial_\beta h^{\alpha \beta}) = 0~.
\end{align}
The above equation describing the evolution of the gravitational perturbation $h_{\mu \nu}$ can be reduced to a familiar form by introducing the following second-rank tensor, 
\begin{equation}\label{17'}
\gamma_{\mu \nu}\equiv h_{\mu \nu}-\frac{1}{2}h\eta_{\mu \nu}-F''(0) \Phi \eta_{\mu \nu}~,
\end{equation}
where $\Phi$ is an extra degree of freedom, to be fixed later. The trace of the above equation defining the tensor $\gamma_{\mu \nu}$ yields, $h=-\gamma -4F''(0)\Phi$. Thus, we can express the perturbation quantity $h_{\mu \nu}$ and its trace, appearing in \ref{18'}, in terms of the tensor $\gamma_{\mu \nu}$ and its trace $\gamma$, along with the scalar degree of freedom $\Phi$, such that \ref{18'} yields, 
\begin{align}
-\frac{1}{2}&\Box \gamma_{\mu \nu}+\frac{1}{2}\partial_{\mu}\partial_\alpha \gamma ^{\alpha}_{\nu} + \frac{1}{2}\partial_\nu \partial_\alpha \gamma^\alpha_\mu 
-F''(0)\eta_{\mu \nu}\Box \Phi
\nonumber 
\\ 
&- \frac{1}{2}\eta_{\mu \nu}(\partial_\alpha \partial_\beta \gamma^{\alpha \beta}) 
+ F''(0)\partial_\mu \partial_\nu \Phi 
\nonumber 
\\ 
& -F''(0)\partial_\mu \partial_\nu R + F''(0)\eta_{\mu \nu} \Box R = 0 ~.
\end{align}
We can impose the usual Lorentz gauge condition on the perturbation variable $\gamma_{\mu \nu}$, i.e., we set $\partial_\alpha \gamma^\alpha _\nu=0$, as well as we identify, $\Phi=R$, such that the field equation for $\gamma_{\mu \nu}$ simplifies to, 
\begin{equation}\label{23'}
\Box \gamma_{\mu \nu}=0~,
\end{equation}
as well as the scalar degree of freedom satisfies the following equation,
\begin{equation}
\label{24'}
    \Box \Phi = \frac{\Phi}{3F''(0)}
\end{equation}
The identification of $\Phi$ with $R$ also provides us the following expression: $\Phi=-\Box h +\partial_\alpha \partial _\beta h^{\alpha \beta}$. Upon expressing $h_{\mu \nu}$ in terms of $\gamma_{\mu \nu}$, $\gamma$ and $\Phi$ using \ref{17'} and using the Lorentz gauge condition, we get back \ref{24'}, confirming the consistency.

Thus the interaction Hamiltonian in the Jordan frame in terms of the new perturbation variables $(\gamma_{\mu \nu},\Phi)$ becomes 
\begin{align}
\label{19a}
\widetilde{H}_{\mathrm{int}}&=-\frac{1}{2}\int d^3\mathbf{r} \; h^{\mu \nu}T_{\mu \nu} 
\nonumber 
\\
&=-\frac{1}{2}\int d^3\mathbf{r}\;\left[\gamma^{\mu \nu}T_{\mu \nu}-\frac{1}{2}\gamma T-2\alpha \Phi T\right]~.
\end{align}
Thus the interaction Hamiltonian between two masses will be different in the $f(R)$ gravity. In the Einstein frame, on the other hand, the total interaction will be,
\begin{equation}
\label{21}
H_{\mathrm{int}}=\widetilde{H}_{\mathrm{int}}+\frac{1}{2}\int d^3\mathbf{r}\;(\Omega^2-1)T~,
\end{equation}
where $\Omega^2=f'(R)=1+2\alpha \Phi$. Thus we can express the interaction Hamiltonian in the Einstein frame as,
\begin{eqnarray}
\label{22}
H_{\mathrm{int}}=-\frac{1}{2}\int d^3 \mathbf{r}\;\left[\gamma^{\mu \nu}T_{\mu \nu}-\frac{1}{2}\gamma T\right]~.
\end{eqnarray}
Thus there is a clear difference between the interaction Hamiltonian between gravity and matter in the Einstein and the Jordan frame, which we hope to persist even in the quantum sector.

\section{Quantization of gravity in Einstein frame}\label{Appquant}

Here, we are interested in linearized gravity around a Minkowski background and its quantization. Such studies have been performed in \cite{Gupta, Suraj_N_Gupta_1950, Donoghue:1994dn} and here we will summarize the basics. Following our conventions, we will take: $\eta_{\mu\nu}=\text{diag}\left(-1,1,1,1\right)$ with $\mu,\nu\in\{0,1,2,3\}$. The perturbed metric $h_{\mu\nu}$ is related to the full metric $g_{\mu\nu}$ by:
\begin{align}
h_{\mu\nu}=\frac{1}{\kappa}\left(g_{\mu\nu}-\eta_{\mu\nu}\right)\,,
\end{align}
where  $\kappa:=\sqrt{16\pi G}=\sqrt{16\pi/M_{\text p}^2}$ and $M_{\text p}$ is the Planck mass. For notational convenience, we will use natural units, i.e., we will set $c=\hbar=1$. Therefore, the total gravitational action becomes,
\begin{align}\label{eq:action_total}
S=S_{\text G}+S_{\text m}+S_{\text{GF}}\,,
\end{align}
where, $S_{\text G}$, $S_{\text m}$, and $S_{\text{GF}}$ are the gravitational, matter, and gauge-fixing actions, respectively. Among these three, the gravitational action $S_{\text G}$ is given by:
\begin{align}
S_{\text G}=\frac14\int\text{d}^4x\big[h^{\mu\nu}\Box h_{\mu\nu}-h\Box h\nonumber\\
+2h^{\mu\nu}\partial_\mu\partial_\nu h-2h^\mu\phantom{}_\alpha\partial_\mu\partial_\nu h^{\nu\alpha}\big]\,,
\label{SG}
\end{align}
The matter action $S_{\text m}$, captures the interaction term between the matter and the perturbed metric, which is given by:
\begin{align}\label{eq:matter_action}
S_{\text int}=-\frac\kappa2\int\text{d}^4x\,h^{\mu\nu}T_{\mu\nu}\,,
\end{align}
where $T_{\mu\nu}$ is the matter energy-momentum tensor. Finally, the gauge-fixing action $S_{\text{GF}}$ is given by:
\begin{align}\label{eq:gauge_fix_action}
S_{\text{GF}}=-\frac12\int\text{d}^4x&\left(\partial_\mu h^\mu\phantom{}_\nu-\frac12\partial_\nu h\right)\nonumber\\
&\times \left(\partial_\alpha h^{\alpha\nu}-\frac12\partial^\nu h\right)\,,
\end{align}
which fixes the gauge freedom in the gravitational action by choosing the Harmonic or the de Donder gauge. Substituting all of these equations, namely~\ref{SG}, \ref{eq:matter_action} and~\ref{eq:gauge_fix_action} into the total action, presented in \ref{eq:action_total}, barring the terms which arise from the total derivatives, we obtain,
\begin{align}\label{eq:action_total_2}
S=\frac14\int\text{d}^4x\bigg[h^{\mu\nu}\Box\left(h_{\mu\nu}-\frac12h\eta_{\mu\nu}\right)-2\kappa h^{\mu\nu}T_{\mu\nu}\bigg]\,.
\end{align}
Following~\cite{Gupta}, it is useful to introduce the following redefined field
\begin{align}
\gamma_{\mu\nu}:=h_{\mu\nu}-\frac12h\eta_{\mu\nu}\,,
\end{align}
and hence the total action in \ref{eq:action_total_2} can be expressed as:
\begin{align}\label{eq:action_with_interaction_redef} 
S=-\frac14&\int\text{d}^4x\bigg[\partial_\alpha\gamma^{\mu\nu}\partial^\alpha\gamma_{\mu\nu}-\frac12\partial^\mu\gamma\partial_\mu\gamma\nonumber\\
&+2\kappa\left(\gamma^{\mu\nu}-\frac12\eta^{\mu\nu}\gamma\right)T_{\mu\nu}\bigg]\,.
\end{align}
We now impose the standard commutation relations between the dynamical variables and their conjugate momentum, which yield \cite{Gupta}:
\begin{align}
\left[\gamma_{\mu\nu}(x),\gamma_{\alpha\beta}(x')\right]&=i\left(\eta_{\mu\alpha}\eta_{\nu\beta}+\eta_{\mu\beta}\eta_{\nu\alpha}\right)D\left(x-x'\right)\,,\label{eq:commutation_relation_combined}\\
\left[\gamma(x),\gamma(x')\right]&=-4iD(x-x')\,,\label{eq:commutation_schwinger_gamma}
\end{align}
Subsequently, both the perturbing fields, namely $\gamma_{\mu\nu}$ and $\gamma$ can be expanded in terms of Fourier modes, such that,
\begin{align}\label{eq:gamma_mu_nu_Fourier}
\gamma_{\mu\nu}=\frac{1}{(2\pi)^{3/2}}\int\frac{\text{d}^3k}{\sqrt{2\omega_{\bm k}}}\left[a_{\mu\nu}\left(\bm k\right){\text e}^{ikx}+a^\dagger_{\mu\nu}\left(\bm k\right){\text e}^{-ikx}\right]\,,
\end{align}
and
\begin{align}\label{eq:gamma_Fourier}
\gamma=\frac{2}{(2\pi)^{3/2}}\int\frac{\text{d}^3k}{\sqrt{2\omega_{\bm k}}}\left[b\left(\bm k\right){\text e}^{ikx}+b^\dagger\left(\bm k\right){\text e}^{-ikx}\right]\,.
\end{align}
In the Fourier space, the commutation relations between the dynamical fields and their conjugate momentum can be expressed as:
\begin{align}
\left[a_{\mu\nu}(\bm k),a^\dagger_{\alpha\beta}(\bm k')\right]&=\left(\eta_{\mu\alpha}\eta_{\nu\beta}+\eta_{\mu\beta}\eta_{\nu\alpha}\right)\delta^{(3)}(\bm k-\bm k')\,,\label{eq:commutation_a_mu_nu}\\
\left[b(\bm k),b^\dagger(\bm k')\right]&=-\delta^{(3)}\left(\bm k-\bm k'\right)\,.\label{eq:commutation_a}
\end{align}
Given the gravitational part of the Lagrangian, along with the dynamical fields and their conjugate momentum,  the gravitational part of the Hamiltonian can be expressed as \cite{Gupta}:
\begin{align}\label{eq:vacuum_hamiltonian}
H_0=\int\text{d}^3k\omega_{\bm k}\left[\frac12a^\dagger_{\mu\nu}(\bm k)a^{\mu\nu}(\bm k)-b^\dagger(\bm k)b(\bm k)\right]\,.
\end{align}
It is worth pointing out that there is a negative sign in the right-hand sides of the commutation relations in \ref{eq:commutation_a_mu_nu} and \ref{eq:commutation_a} for $a_{0i}\left(\bm k\right)$ and $b\left(\bm k\right)$, respectively. In our convention, the Roman indices run as $i\in\{1,2,3\}$. Therefore, the eigenvalues of the operators $\int\text{d}^3\bm{k}~\omega_{\bm k}a_{0i}^\dagger\left(\bm k\right)a_{0i}\left(\bm k\right)$ and $\int\text{d}^3\bm{k}~\omega_{\bm k}b^\dagger\left(\bm k\right)b\left(\bm k\right)$ are positive \cite{Gupta,Suraj_N_Gupta_1950}. 

Thus, when the Hamiltonian acts on some state in the Hilbert space, the terms containing $a_{0i}\left(\bm k\right)$ and $b\left(\bm k\right)$ operators contribute positively to the energy, since their coefficients are negative in \ref{eq:vacuum_hamiltonian}. Therefore, it follows that the energy values associated with the Hamiltonian in \ref{eq:vacuum_hamiltonian}, acting on some general states are positive. Nevertheless, there is still the issue of whether these states can have negative probabilities. To get rid of these negative probability states, one imposes supplementary conditions \cite{Gupta, Suraj_N_Gupta_1950, Vinckers:2023grv} leading to physical states having only two polarizations and positive probabilities; thus ensuring that the gravitational Hamiltonian is bounded from below. 

We will now discuss these supplementary conditions, and for that purpose define the general state $\ket{\psi}$ as the sum over $n_{\mu\nu}$ $a_{\mu\nu}\left(\bm k\right)$ and $n$ $b\left(\bm k\right)$. As discussed earlier, the action of the Hamiltonian, described by \ref{eq:vacuum_hamiltonian}, on such a general state yields positive energy values. However, in order for the Hamiltonian to be bounded from below, it is necessary to introduce the following supplementary conditions \cite{Gupta}, 
\begin{align}\label{eq:supplementary_condition_1}
\partial^\mu\gamma_{\mu\nu}^{(+)}\ket{\psi}=0\,,
\end{align}
where $\gamma_{\mu\nu}^{(+)}$ refers to the ${\text e}^{ikx}$ part in \ref{eq:gamma_mu_nu_Fourier}. Without loss of generality, we can align $\bm k$ along the $z$-axis, and hence the supplementary condition above yields the following constraints
\begin{align}
\left[a_{00}\left(\bm k\right)-a_{03}\left(\bm k\right)\right]\ket{\psi}=0\,,\label{eq:supplementary_a_00}\\
\left[a_{01}\left(\bm k\right)-a_{13}\left(\bm k\right)\right]\ket{\psi}=0\,,\label{eq:supplementary_a_01}\\
\left[a_{02}\left(\bm k\right)-a_{23}\left(\bm k\right)\right]\ket{\psi}=0\,,\label{eq:supplementary_a_02}\\
\left[a_{03}\left(\bm k\right)-a_{33}\left(\bm k\right)\right]\ket{\psi}=0\,.\label{eq:supplementary_a_03}
\end{align}
Let us first consider \ref{eq:supplementary_a_00}, from which, we note that $a_{00}\left(\bm k\right)$ satisfies a positive commutation relation, while $a_{03}\left(\bm k\right)$ satisfies a negative one. Thus it follows that any state satisfying \ref{eq:supplementary_a_00} and involving $a_{00}\left(\bm k\right)$ or $a_{03}\left(\bm k\right)$ will be redundant, implying the action of the Hamiltonian on such a state will identically vanish. Therefore, we may set $n_{00}=n_{03}=0$; reducing the number of polarizations from eleven to nine. 
 
Next, we study the supplementary condition given by \ref{eq:supplementary_a_01}. From the commutation relation \ref{eq:commutation_a_mu_nu} it follows that $a_{01}\left(\bm k\right)$ satisfies a negative commutation relation, while $a_{13}\left(\bm k\right)$ satisfies a positive one. Thus we may set $n_{01}=n_{13}=0$, since any $a_{01}\left(\bm k\right)$ or $a_{13}\left(\bm k\right)$ excitations would lead to the redundant contributions. \ref{eq:supplementary_a_01} thus reduces the number of polarizations from nine to seven. 
 
The same argument also applies to \ref{eq:supplementary_a_02} and \ref{eq:supplementary_a_03}. The first constraint yields $n_{02}=n_{23}=0$ and the number of polarizations reduces from seven to five. While, from \ref{eq:supplementary_a_03} we obtain $n_{03}=n_{33}=0$. However, we already have $n_{03}=0$ as a result of \ref{eq:supplementary_a_00}. Therefore, the condition \ref{eq:supplementary_a_03} reduces the number of polarizations by a single degree of freedom, i.e., from five to four.

The second supplementary condition, also imposed in \cite{Gupta}, takes the following form, 
\begin{align}\label{eq:supplementary_condition_2}
\left[\gamma_\mu^{(+)}\phantom{}^\mu-\gamma^{(+)}\right]\ket{\psi}=0\,,
\end{align}
where, as in the earlier case, $\gamma^{(+)}$ refers to only the $\text{\ e}^{ikx}$ part in \ref{eq:gamma_Fourier}. By defining,
\begin{align}
\sqrt2a'_{11}:=a_{11}-a_{22}\,,\ \ \ \ \ \ \ \ \sqrt2a'_{22}:=a_{11}+a_{22}\,,
\end{align}
and making use of the fact that there are no $a_{00}\left(\bm k\right)$ or $a_{33}\left(\bm k\right)$ states, thanks to the first supplementary condition in \ref{eq:supplementary_condition_1}, the second supplementary condition yields
\begin{align}
\left[\sqrt2a'_{22}\left(\bm k\right)-b\left(\bm k\right)\right]\ket{\psi}=0\,.
\end{align}
In this case the operator $a'_{22}\left(\bm k\right)$ satisfies a positive commutation relation, while $b\left(\bm k\right)$ satisfies a negative one. Therefore, it follows that $a'_{22}\left(\bm k\right)$ or $b\left(\bm k\right)$ do not contribute to excited states. The number of polarizations for the physical state $\ket{\psi}$ is now reduced from four to two through the use of the second supplementary condition; described by $a'_{11}\left(\bm k\right)$ and $a_{12}\left(\bm k\right)$ excitations. 

Finally, since there are no $a_{0i}\left(\bm k\right)$ or $b\left(\bm k\right)$ excitations satisfying negative commutation relations, it follows that there are no physical states with the negative norm. Thus the Hamiltonian, as in \ref{eq:vacuum_hamiltonian}, is bounded from below, provided both of the supplementary conditions, as in \ref{eq:supplementary_condition_1} and \ref{eq:supplementary_condition_2} are satisfied.

Denoting the ground state of the system as $\ket{0}$, the relativistic one-particle normalized states can be defined as,
\begin{align}
\ket{\bm k}_{\mu\nu}:=\sqrt{2\omega_{\bm k}}\left[a^\dagger_{\mu\nu}(\bm k)-\eta_{\mu\nu}b^\dagger(\bm k)\right]\ket{0}\,.
\end{align}
Applying the well-known perturbation theory (see for example \cite{Scadron:1991ep}), the shift in the gravitational energy $\Delta H_{\rm g}$ between two gravitationally interacting point masses, to second order in $\kappa$ yields,
\begin{align}\label{eq:ham_trajectories}
\Delta \hat H=-\frac{\kappa^2}{2}\int\text{d}^3k\frac{\bra{0}V\ket{\bm k}_{\mu\nu}\eta^{\mu\alpha}\eta^{\nu\beta}_{~~~\alpha\beta}{\bra{\bm k}}{}V\ket{0}}{\bm k^2}\,.
\end{align}
As noted in \cite{Bose:2022uxe, Vinckers:2023grv}, there is no ${\cal O}(\kappa)$ correction to the interaction Hamiltonian since it involves inner products of the ground state with the first excited state only. Therefore, the shift in the gravitational energy between two point masses can be computed as in \cite{Bose:2022uxe, Vinckers:2023grv}, and yields:
\begin{align}
\Delta\hat{H}_{g}&=-\frac{Gm^{2}}{\vert{\hat{\bf r}}_{A}-{\hat{\bf r}}_{B}\vert}\nonumber \\
&  -\frac{G(3{\hat{p}}_{A}^{2}-8\hat{p}_{A}\hat{p}_{B}+3{\hat{p}}_{B}^{2})}{2c^{2}\vert{\hat{\bf r}}_{A}-{\hat{\bf r}}_{B}\vert} \nonumber \\
&  + \frac{G(5{\hat{p}}_{A}^{4}-18{\hat{p}}_{A}^{2}{\hat{p}}_{B}^{2}+5{\hat{p}}_{B}^{4})}{8c^{4}m^{2}\vert{\hat{\bf r}}_{A}-{\hat{\bf r}}_{B}\vert}.\label{eq:dH5a}
\end{align}
at the lowest order in the perturbation theory. The result matches the earlier computations performed by many, see~\cite{Donoghue:1994dn, Grignani:2020ahv, Bose:2022uxe, Vinckers:2023grv}. We have used these results in the main text.

\bibliography{References.bib}

\end{document}